\documentclass[aps,prb,showpacs,superscriptaddress, preprint numbers,twocolumn]{revtex4-1}
\usepackage{rotating}
\usepackage{epstopdf}
\usepackage{color}
\usepackage{amsmath}
\usepackage{graphicx}
\usepackage{hyperref}
\usepackage{color}
\usepackage{amssymb}
\usepackage{dcolumn}
\usepackage{bm}
\usepackage{epsfig}
\usepackage{array}
\usepackage{subfigure}
\usepackage{upgreek}
\usepackage{wasysym}
\usepackage{slashed}
\usepackage[section]{placeins}
\usepackage{appendix}
\usepackage{subfigure}
\usepackage{multirow}
\newcommand{\be}{\begin{equation}} 
\newcommand{\ee}{\end{equation}} 
\newcommand{\bea}{\begin{eqnarray}} 
\newcommand{\eea}{\end{eqnarray}} 
\newcommand{\bqa}{\begin{eqnarray}}
\newcommand{\eqa}{\end{eqnarray}}

\newcommand{\figphononA}{
\begin{figure}[htb]
\centering
\hspace{0cm}
\subfigure[noonleline][]
{\label{fig:memory_LA}\includegraphics[height=35mm,width=40mm]{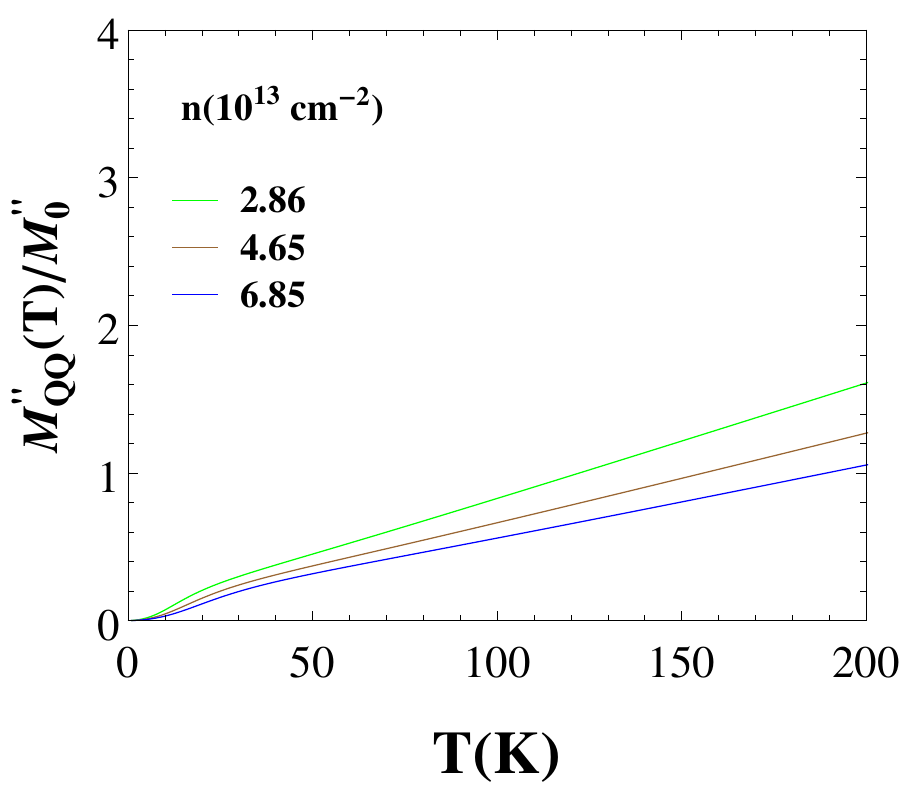}}
\hspace{0cm}
\subfigure[noonleline][]
{\label{fig:memory_TA}\includegraphics[height=35mm,width=40mm]{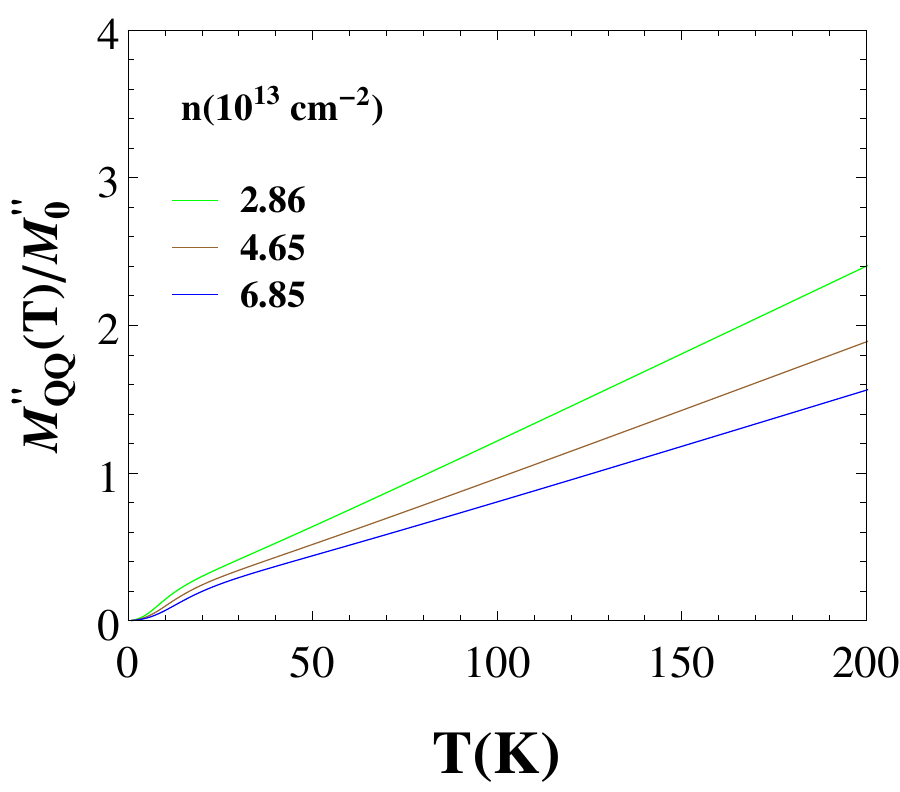}}
\caption{The imaginary part of the thermal memory function for the longitudinal and transverse acoustic phonons are plotted with temperature at different $\Theta_{\text{BG}} \propto \sqrt{n}$. (a): for the longitudinal acoustic (LA) phonons and (b): for the transverse acoustic (TA) phonons.}
\label{fig:memory_LATA}
\end{figure}
}
\newcommand{\figphononB}{
\begin{figure}[htb]
\centering
\hspace{0cm}
\subfigure[noonleline][]
{\label{fig:thermal_LA}\includegraphics[height=35mm,width=40mm]{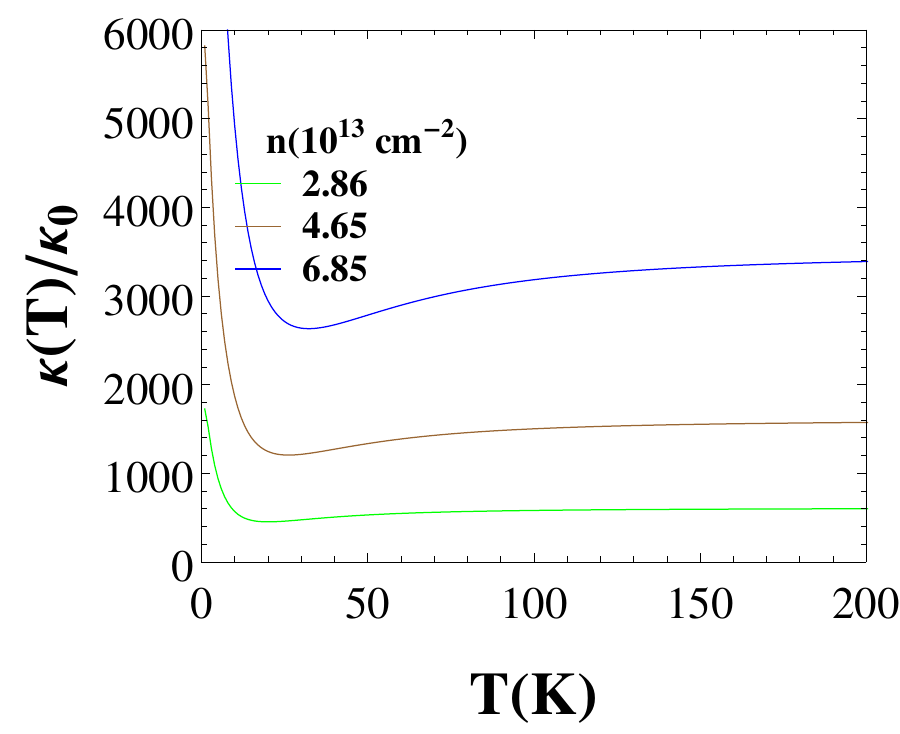}}
\hspace{0cm}
\subfigure[noonleline][]
{\label{fig:thermal_TA}\includegraphics[height=35mm,width=40mm]{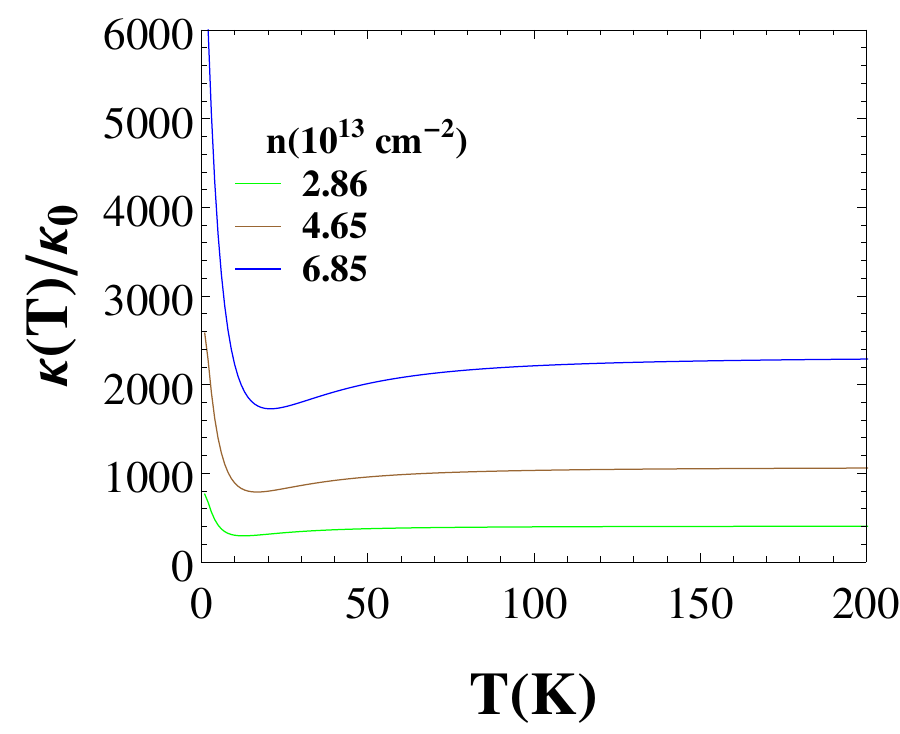}}
\caption{The Thermal Conductivity for the longitudinal and transverse acoustic phonons are plotted with temperature at different $\Theta_{\text{BG}} \propto \sqrt{n}$. (a): for the longitudinal acoustic (LA) phonons and (b): for the transverse acoustic (TA) phonons.}
\label{fig:thermal_LATA}
\end{figure}
}
\newcommand{\figphononC}{
\begin{figure}[htb]
\centering
\hspace{0cm}
\subfigure[noonleline][]
{\label{fig:memory_ac_LA}\includegraphics[height=35mm,width=40mm]{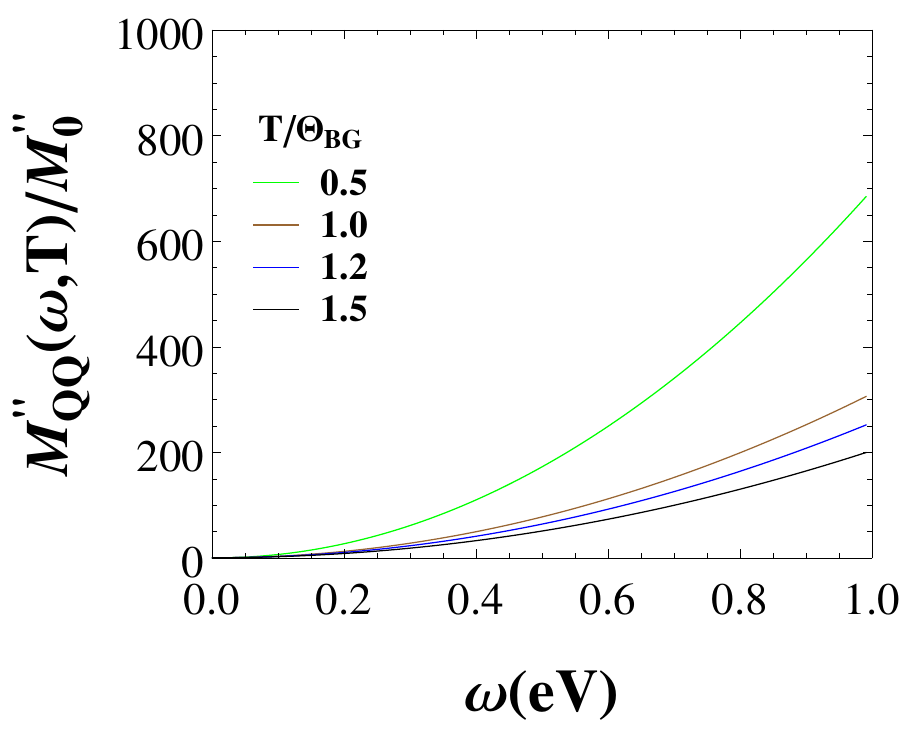}}
\hspace{0cm}
\subfigure[noonleline][]
{\label{fig:memory_ac_TA}\includegraphics[height=35mm,width=40mm]{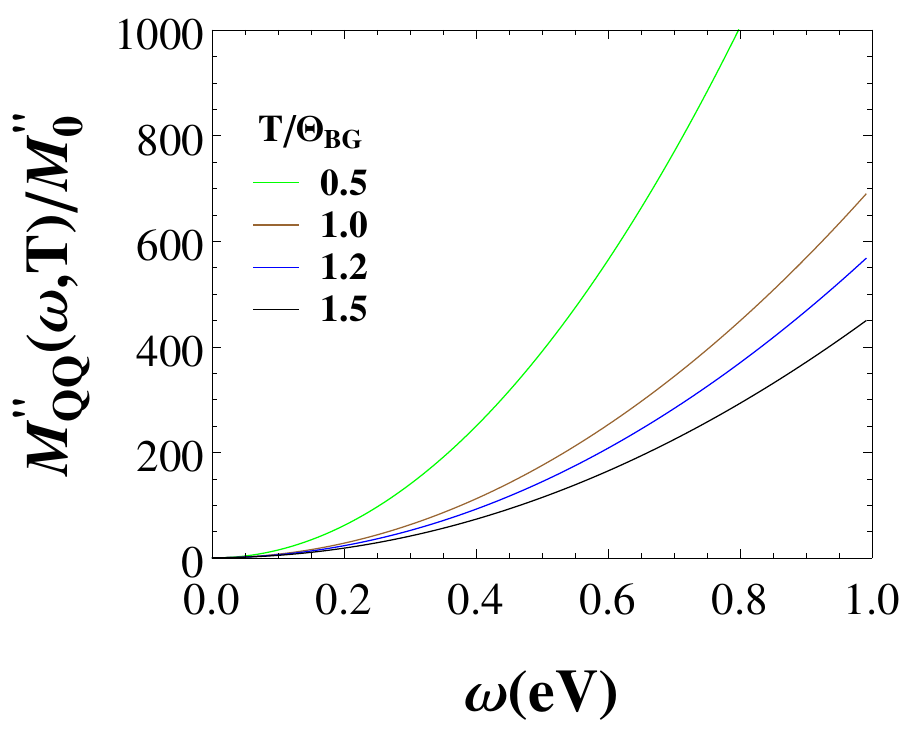}}
\caption{The frequency and temperature dependent thermal memory function or the thermal scattering rate for the longitudinal and transverse acoustic phonons are plotted with frequency at different $T/\Theta_{\text{BG}}$ ratio. (a): for the longitudinal acoustic (LA) phonons and (b): for the transverse acoustic (TA) phonons.}
\label{fig:memory_ac_LATA}
\end{figure}
}
\newcommand{\figphononD}{
\begin{figure}[htb]
\centering
\hspace{0cm}
\subfigure[noonleline][]
{\label{fig:memory_ac_LA_low}\includegraphics[height=35mm,width=40mm]{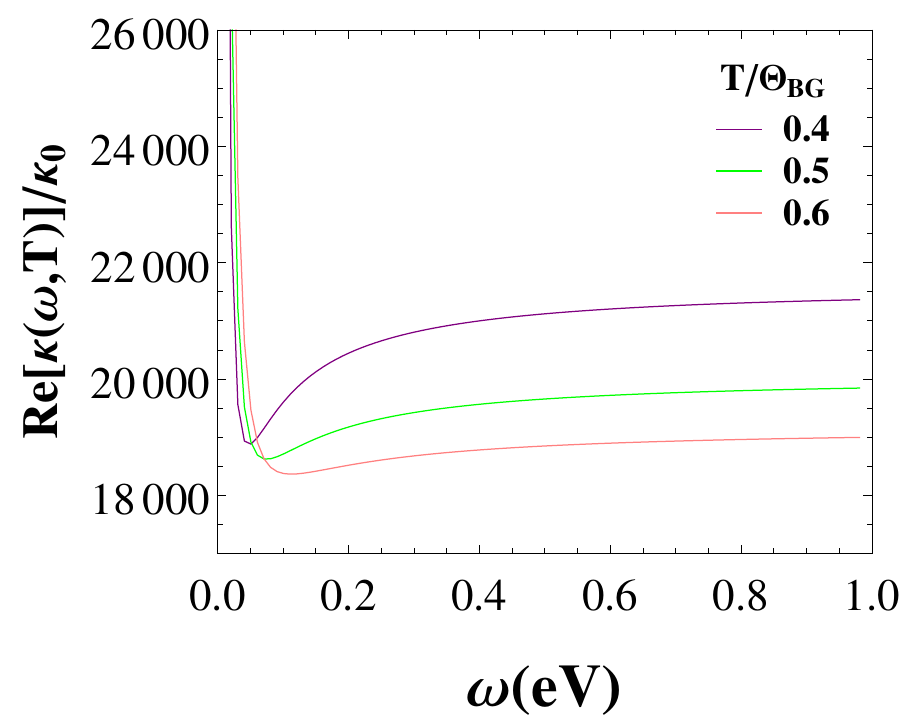}}
\hspace{0cm}
\subfigure[noonleline][]
{\label{fig:memory_ac_LA_high}\includegraphics[height=35mm,width=40mm]{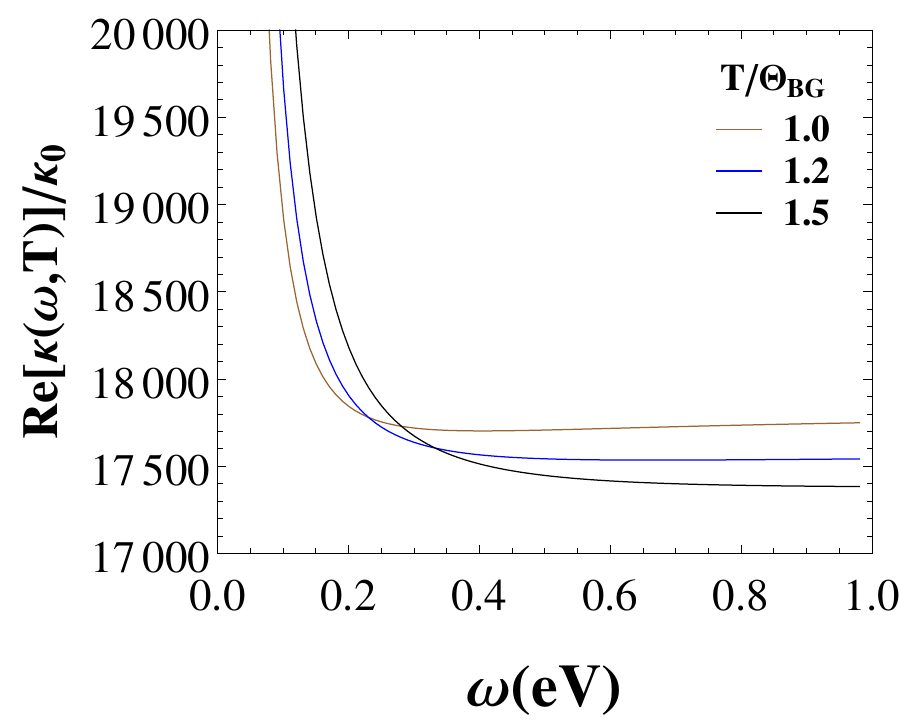}}
\caption{The frequency and temperature dependent thermal conductivity for the longitudinal acoustic phonons is plotted with frequency at different $T/\Theta_{\text{BG}}$ ratio.}
\label{fig:thermal_ac_LA}
\end{figure}
}
\newcommand{\figphononF}{
\begin{figure}[htb]
\centering
\hspace{0cm}
\subfigure[noonleline][]
{\label{fig:memory_ac_TA_low}\includegraphics[height=35mm,width=40mm]{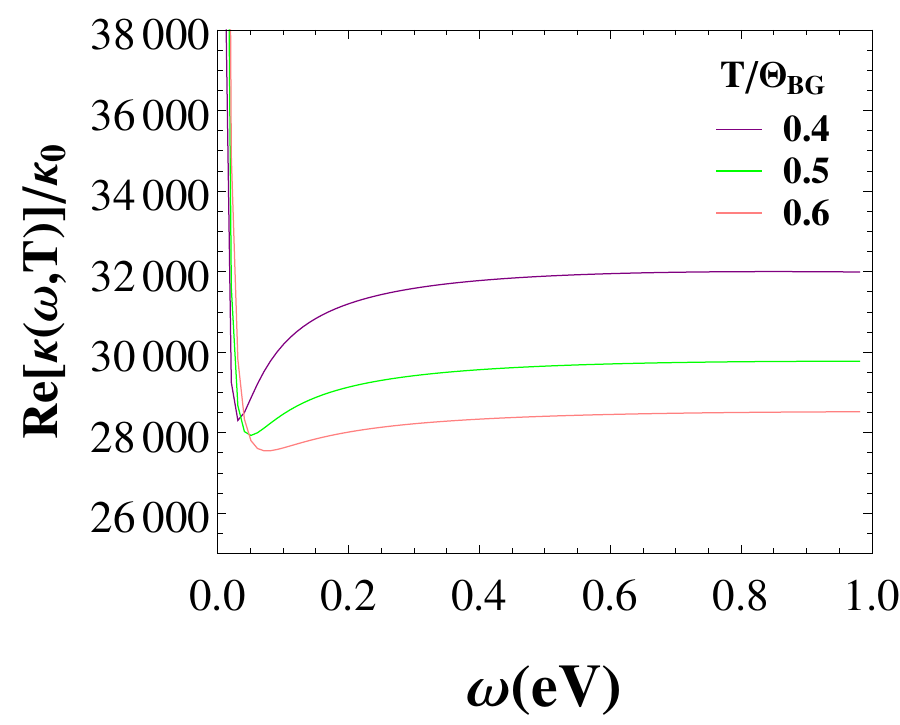}}
\hspace{0cm}
\subfigure[noonleline][]
{\label{fig:memory_ac_TA_high}\includegraphics[height=35mm,width=40mm]{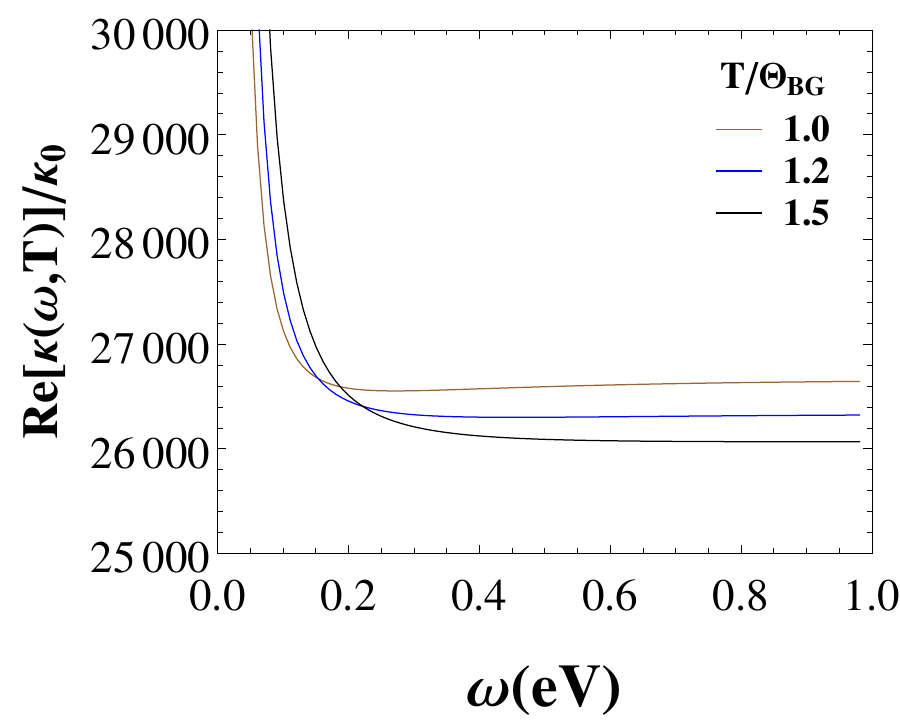}}
\caption{The frequency and temperature dependent thermal conductivity for the transverse acoustic phonons is plotted with frequency at different $T/\Theta_{\text{BG}}$ ratio.}
\label{fig:thermal_ac_TA}
\end{figure}
}
\newcommand{\figphononE}{
\begin{figure}[htb]
\centering
\hspace{0cm}
\subfigure[noonleline][]
{\label{fig:memory_ZA}\includegraphics[height=35mm,width=40mm]{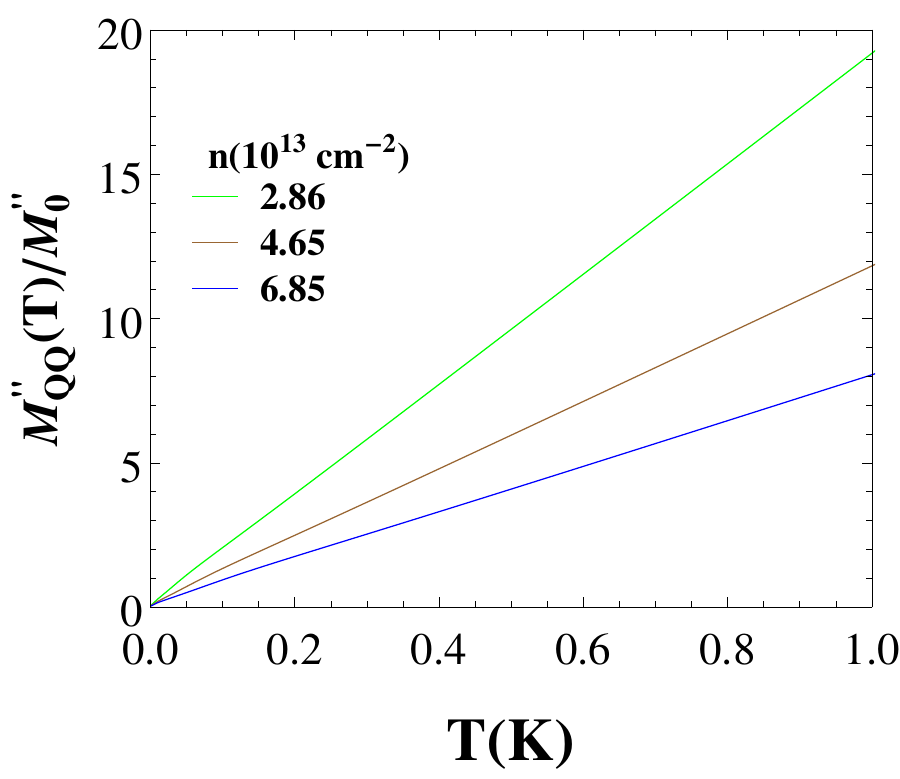}}
\hspace{0cm}
\subfigure[noonleline][]
{\label{fig:thermal_ZA}\includegraphics[height=35mm,width=40mm]{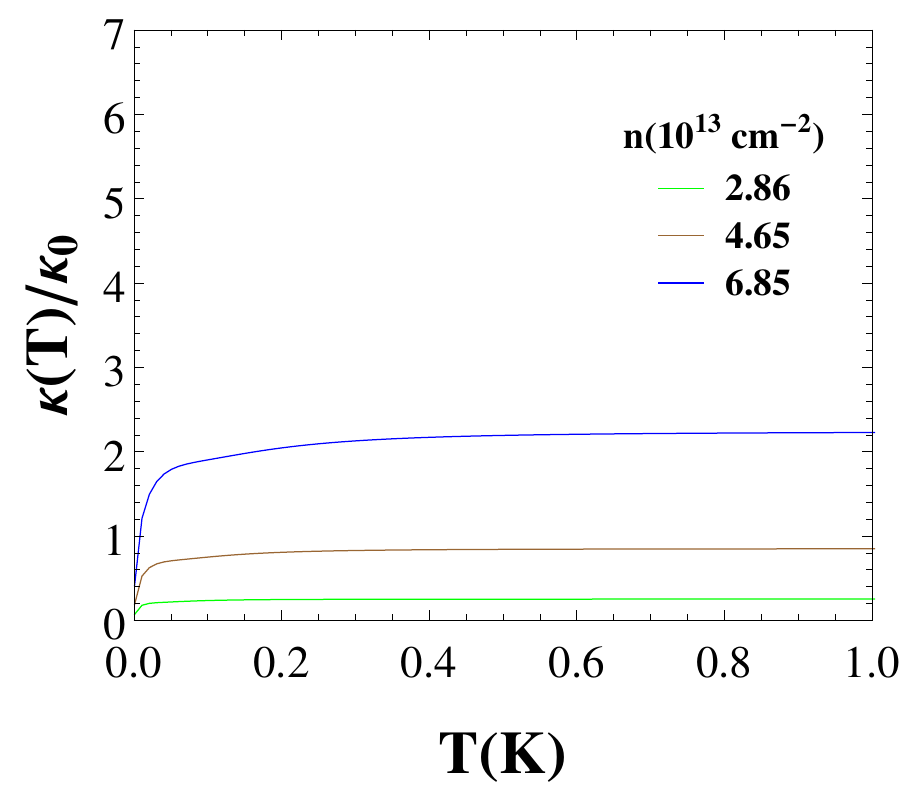}}
\caption{(a): The Thermal memory function for the flexural acoustic phonons (ZA) is plotted with temperature at different $\Theta_{\text{BG}} \propto n$ and (b): the corresponding thermal conductivity.}
\label{fig:thermal_memory_ZA}
\end{figure}
}
\newcommand{\figphononG}{
\begin{figure}[htb]
\centering
\hspace{0cm}
\subfigure[noonleline][]
{\label{fig:memory_ac_ZA}\includegraphics[height=35mm,width=40mm]{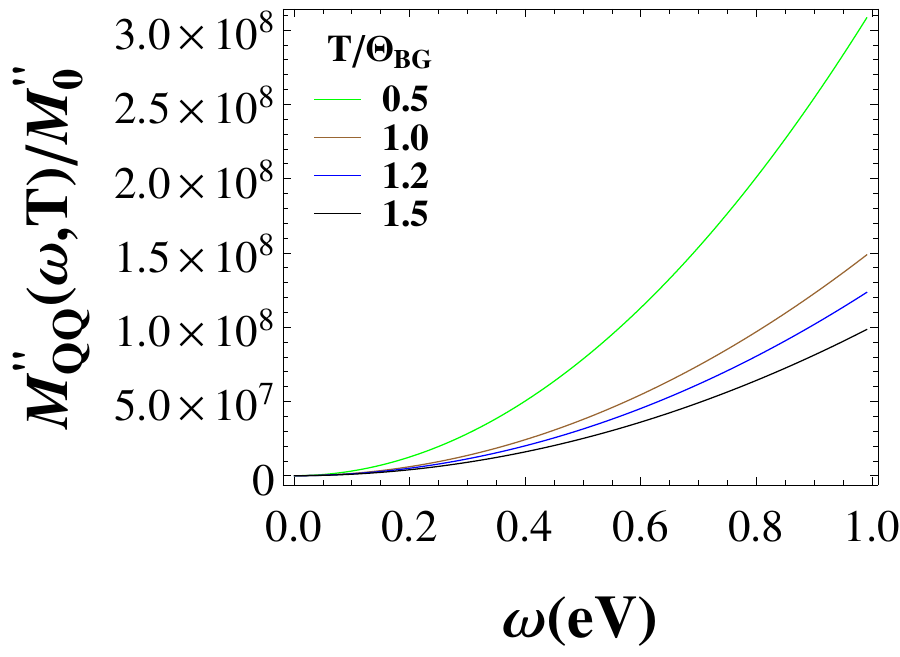}}
\hspace{0cm}
\subfigure[noonleline][]
{\label{fig:thermal_ac_ZA}\includegraphics[height=35mm,width=40mm]{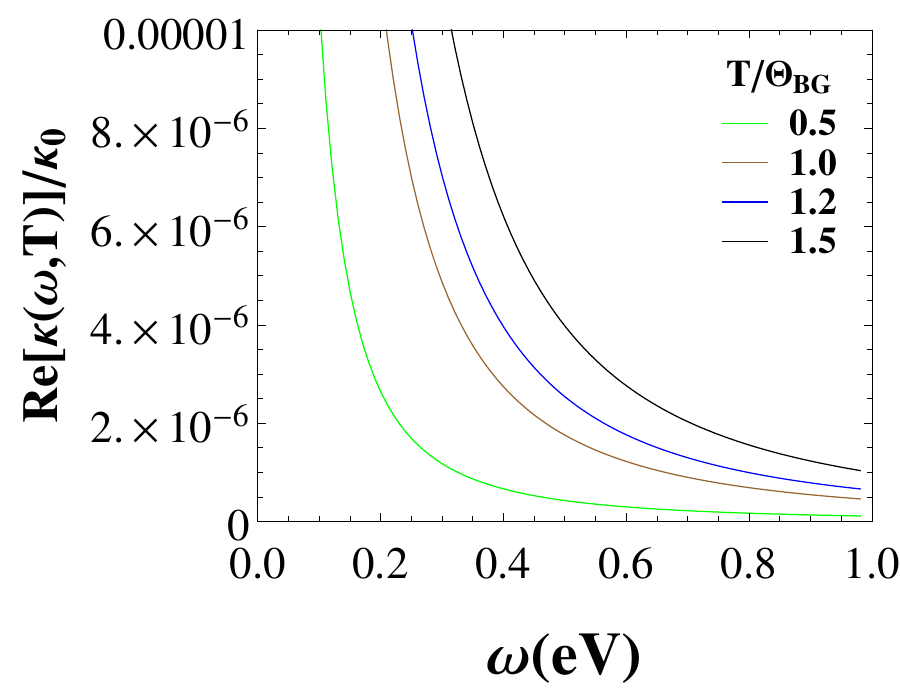}}
\caption{(a): The Thermal memory function for the flexural acoustic phonons (ZA) is plotted with frequency $T/\Theta_{\text{BG}}$ ratio. (b): The corresponding Thermal Conductivity variation of flexural phonons.}
\label{fig:thermal_memory_ac_ZA}
\end{figure}
}
\begin{document}
\title{Role of acoustic phonons in frequency dependent thermal conductivity of graphene}

\author{Pankaj Bhalla}
\email{pankajbhalla66@gmail.com}
\affiliation{Physical Research Laboratory, Navrangpura, Ahmedabad-380009 India.}
\affiliation{Indian Institute of Technology Gandhinagar-382424, India.}

\date{\today}
\begin{abstract}
We study the effect of the electron-phonon interaction on the finite frequency dependent electronic thermal conductivity of two dimensional graphene. We calculate it for various acoustic phonons present in graphene and characterized by different dispersion relations using the memory function approach. It is found that the thermal conductivity $\kappa(T)$ in the zero frequency limit follows different power law for the longitudinal/transverse and the flexural acoustic phonons. For the longitudinal/transverse phonons, $\kappa(T) \sim T^{-1}$ at the low temperature and saturates at the high temperature. These signatures are qualitatively agree with the results predicted by the Boltzmann equation. Similarly, for the flexural phonons, we find that $\kappa(T)$ shows $T^{1/2}$ law at the low temperature and then saturates at the high temperature. In the finite frequency regime, we observe that the real part of the thermal conductivity, $\text{Re}[\kappa(\omega,T)]$ follows $\omega^{-2}$ behavior at the low frequency and becomes frequency independent at the high frequency.
\end{abstract}
\maketitle

\section{Introduction}
In recent times, Graphene\cite{geim_07, sarma_07, geim_09} has attracted a lot of attention both in 
the fundamental and applied research due to its unique electronic and optical properties. These properties include anomalous high electrical conductivity, high thermal conductivity, quantum Hall effect, effect of impurities on the electric properties, etc.\cite{sarma_07a, sarma_07b, sarma_07c, sarma_07d, sarma_08b, sarma_08a, sarma_09a, sarma_09b, sarma_09, peres_10, abergel_10, cooper_12} which make the use of this material quite promising for the fabrication or design of the electronic devices. Among these properties, electrical conductivity, Hall effect have been discussed several times in literature, while there is lack of discussions in the electronic contribution to the thermal conductivity. Thus, in the present work, we focus on the thermal conductivity of graphene.

In the literature, it is argued that the unusual high thermal conductivity of graphene\cite{baladin_08, nika_12} is mainly contributed by the phonons and the electronic contribution is small, hence neglected. 
However, in real systems, the total thermal conductivity is expressed as the sum of the electronic and phononic thermal conductivity which are different in different temperature regimes. In the low temperature limit, the thermal conductivity mainly arises due to the scattering of electrons or phonons by impurities. As the temperature increases, the number of phonons increases which further enhance the electron-phonon and phonon-phonon scatterings. It has been depicted by the Boltzmann approach that at the high temperature i.e. $T \gg \Theta_{D}$, $\Theta_{D}$ being the Debye temperature, the thermal conductivity shows temperature independent behavior\cite{wilson_book, ziman_book, bhalla_16}. While as the temperature decreases below $\Theta_{D}$, only the acoustic phonons within the phonon sphere of radius $k_{\text{ph}}$ with $k_{\text{ph}} \ll k_{D}$, where $k_{D}$ is the radius of Debye sphere, play a role in the thermal conductivity\cite{wilson_book, ziman_book, bhalla_16}. In the three dimensional metals, it leads to $T^{-2}$ behavior of the thermal conductivity. In such systems, the radius of the Fermi sphere is larger than the radius of the Debye sphere i.e. $2k_{F} \gg k_{D}$. Thus all phonons can scatter off the electrons. But in the systems where $k_{F} \ll k_{D}$, only small number of phonons can  scatter off the electrons. These phonons are restricted within the energy range $v_{s}k_{\text{ph}} \leq 2v_{s}k_{F}$. This can be explained by introducing the new temperature scale known as Bloch Gr\"uneisen (BG) temperature which is smaller than the Debye temperature\cite{efetov_10}. This scale defines two regimes i.e. low temperature ($T \ll \Theta_{\text{BG}}$) and high temperature ($T \gg \Theta_{\text{BG}}$) regimes for the electron-phonon interaction in graphene. In the low temperature regime ($T \ll \Theta_{\text{BG}}$), the acoustic phonons with linear dispersion relation yield inverse temperature behavior to the thermal conductivity (i.e. $\kappa \sim T^{-1}$) and then change to the temperature independent behavior in the high temperature regime ($T \gg \Theta_{\text{BG}})$\cite{munoz_12, kim_16}. However, because of hexagonal crystal structure of the graphene, there are also other acoustic phonons known as flexural phonons or out of plane phonons which obey quadratic dispersion relation and hence give different power law behavior to the electronic thermal conductivity. Thus the role of the different acoustic phonons is very important to understand the transport or the thermal conductivity of graphene. However most of the studies has considered only the zero frequency limit. But for the generation of the integrated circuits, high frequency communication devices, the study of the thermal conductivity in the dynamical regime is important as it may degrade the issue of the heat dissipation within the systems\cite{volz_01, shastry_06, koh_07, shastry_09}.

With this motivation, we have examined the electronic thermal conductivity both in the zero frequency and the finite frequency regime using the memory function approach\cite{zwanzig_61, zwanzig_61a, mori_65, forster_95, fulde_12}. The advantage of using memory function approach is that it directly deals with the dynamics of the transport\cite{nabyendu_16}. Here we discuss the dynamical behavior of the thermal conductivity due to the interactions of electrons with different acoustic phonons and also its difference with the behavior in normal metals. In the zero frequency limit, our findings for the thermal conductivity of graphene agrees qualitatively with the results predicted by the Boltzmann approach\cite{munoz_12, kim_16}. In the finite frequency regime, our findings may be important from both the fundamental and the application point of view and may inspire important experimental studies in future.

This paper is organized as follows. In Sec.\ref{sec:theroretical_framework}, first we discuss the basic idea of the thermal conductivity and its relation with the memory function. Then the model Hamiltonian considering only the electron-phonon interactions in graphene is discussed. Later, we discuss the phonon dispersion relation of different acoustic phonons. With these descriptions, we calculate the finite frequency and temperature dependent thermal conductivity for different acoustic phonons. In Sec.\ref{sec:results}, the results are presented in the two subsections. In one subsection, we discuss the thermal conductivity in the zero frequency limit. In other subsection, the results for the finite frequency in different BG regimes has been discussed. Finally, in Sec.\ref{sec:conclusion}, we conclude.

\section{Theoretical Framework}
\label{sec:theroretical_framework}
\subsection{Thermal Conductivity}
\label{subsec:thermal_conductivity}
According to the Kinetic theory, the thermal conductivity is defined as the rate of flow of heat across a unit area of cross section in a unit temperature gradient\cite{mahan_book} i.e.
\bea
J_{Q} &=& -\kappa \nabla T.
\label{eqn:thermal_currentdefinition}
\eea
Here $J_{Q}$ is the thermal current density and is defined as,
\bea
J_{Q} = \frac{1}{m} \sum_{\textbf{k}} \textbf{k}.\hat{n} (\epsilon_{\textbf{k}} - \mu) c_{\textbf{k}}^{\dagger} c_{\textbf{k}},
\label{eqn:thermal_current}
\eea
where $c_{\textbf{k}} (c_{\textbf{k}}^{\dagger})$ is the annihilation(creation) operator having momentum $\textbf{k}$, $\epsilon_{\textbf{k}}$ is the electron energy dispersion of graphene, $\mu$ is the chemical potential, $m$ is the electron mass and $\hat{n}$ is the unit vector parallel to the direction of heat current. And in Eq.(\ref{eqn:thermal_currentdefinition}) $\nabla T$ is the temperature gradient and $\kappa$ is the thermal conductivity. The later is known as response due to the change in the temperature gradient and is generally analyzed by various approaches\cite{wilson_book, ziman_book} where the gradient of the temperature is considered as static. But in the present work, we assume that $\nabla T$ is not static, while it oscillates with the frequency $\omega$. This oscillation leads to the dynamical variation of the thermal conductivity. Here we set $\hbar = 1$ and $k_{B}=1$ in our calculations.\\
To compute it, we employ the memory function approach. Following the later approach, the dynamical thermal conductivity at complex frequency $z$ and temperature $T$ is defined as\cite{bhalla_16}
\bea
\kappa (z,T) &=& \frac{i}{T} \frac{\chi_{QQ}^{0}(T)}{z+ M_{QQ}(z,T)},
\label{eqn:thermal_conductivity}
\eea
where $\chi_{QQ}^{0}(T)$ is the static thermal current-thermal current correlation function i.e. $\chi_{QQ}^{0}(T) = \frac{\pi}{24} \frac{k_{F}^{3}}{m^{2}v_{F}} T^{2}$, where $k_{F}$ is the Fermi wave vector and $v_{F}$ is the Fermi velocity, $M_{QQ}(z,T)$ is the thermal memory function.\\
It is known that within the perturbation theory, the thermal memory function can be expressed to the leading order in the electron-phonon coupling, as\cite{bhalla_16a, nabyendu_16, bhalla_16}
\bea \nonumber &&
M_{QQ}(z,T)\\ &&=\frac{\langle \langle [J_{Q},H]; [J_{Q},H] \rangle \rangle_{z=0} - \langle \langle [J_{Q},H]; [J_{Q},H] \rangle \rangle_{z}}{z\chi_{QQ}^{0}(T)}. 
\label{eqn:thermal_memoryfunction}
\eea
This is the complex memory function in which the imaginary part of the memory function describes the thermal scattering rate and the real part describes the mass enhancement factor. In the present work, we focus on the thermal scattering rate which leads to the real part of the thermal conductivity. Here for simplicity, we have ignored the mass enhancement contribution to the thermal conductivity. To calculate it, we require the total Hamiltonian that is discussed in the next subsection.\\
\subsection{Model Hamiltonian}
\label{subsec:hamiltonian}
We consider a two dimensional graphene with only electron-phonon interactions. Then, the Hamiltonian of such a system is described as
\bea
H &=& H_{0} + H_{\text{ep}} + H_{\text{ph}},
\label{eqn:hamiltonian}
\eea
where $H_{0} = \sum_{\textbf{k} \sigma} \epsilon_{\textbf{k}} c^{\dagger}_{\textbf{k} \sigma} c_{\textbf{k} \sigma}$ and $H_{\text{ph}} = \sum_{q} \omega_{q} \left( b_{q}^{\dagger} b_{q} +\frac{1}{2} \right)$ corresponds to the Hamiltonians of the free electrons and phonons respectively. Here $\omega_{q}$ is the phonon energy dispersion, $b_{q}(b_{q}^{\dagger})$ is the phonon annihilation(creation) operator having phonon wave vector $\textbf{q} = \textbf{k} - \textbf{k}'$ and $\sigma$ is the electron spin. $H_{\text{ep}}$ describes the electron-phonon interactions and is given as  $H_{\text{ep}} = \sum_{\textbf{k} \textbf{k}' \sigma} \left[ D(\textbf{k}-\textbf{k}') c^{\dagger}_{\textbf{k} \sigma} c_{\textbf{k}' \sigma} b_{\textbf{k}-\textbf{k}'} + H.c. \right]$,
where $D(q)$ is the electron-phonon matrix element. The later is usually written in the following form\cite{sarma_08, sarma_11}
\bea
D(\textbf{q}) &=& \frac{D_{0}q}{\sqrt{2 \rho_{m} \omega_{q}}} \left( 1-\left(\frac{q}{2k_{F}}\right)^{2} \right)^{1/2}.
\label{eqn:electron-phononmatrixelement}
\eea
Here $D_{0}$ is the deformation potential coupling constant, $\rho_{m}$ is the graphene mass density and $\omega_{q}$ is the phonon energy dispersion.

\subsection{Phonon Dispersions}
\label{subsec:phonon_dispersion}
Before proceeding to compute the thermal scattering rate and the corresponding thermal conductivity for the sake of completeness, we will first discuss the phonon dispersion relations in this subsection.

The thermal transport due to the electron-phonon interactions significantly depends on the characteristics of the phonon which are further determined by the two dimensional structure of the graphene. In graphene, there are two carbon atoms per hexagonal unit cell which gives six phonon branches in the dispersion spectrum. These are three acoustic and three optical branches namely LA(Longitudinal Acoustic), TA(Transverse Acoustic), LO(Longitudinal Optical), TO(Transverse Optical), ZA(Flexural Acoustic) and ZO(Flexural Optical). The TA and TO phonons are due to the transverse vibrations within the graphene plane and LA, LO are due to the longitudinal vibrations within the graphene plane. The other modes such as ZA, ZO are due to the oscillations of phonons in the direction normal to the longitudinal and transverse phononic modes. These phononic modes are also referred to the out of plane modes\cite{sanders_13}. In the present work, we deal with the low energy excitations, thus only acoustic phonons are considered in throughout the calculations.

From the phonon dispersion spectra, it has been found that these modes follow different dispersion relations. The LA and TA modes follow the linear dispersion relations\cite{sanders_13, kaasbjerg_12} i.e.
\bea \nonumber
\omega_{\text{LA}} &\approx& v_{\text{LA}} q    \\ \omega_{\text{TA}} &\approx& v_{\text{TA}} q,
\eea
where $v_{\text{LA}}$ and $v_{\text{TA}}$ are the longitudinal and transverse phonon velocities and $v_{\text{LA}} = 21.3 \times 10^{3}$ms$^{-1}$, $v_{\text{TA}} = 14.1 \times 10^{3}$ms$^{-1}$.\cite{kaasbjerg_12} 

The other acoustic phonon ZA approximately follows the quadratic dispersion relation as\cite{sanders_13, amorim_13}
\bea
\omega_{\text{flex}} &\approx& \alpha q^{2}.
\eea
Here the parameter $\alpha = \left(\frac{s}{\rho_{m}}\right)^{1/2}$, where $s$ is the bending stiffness of the graphene, $\rho_{m}$ is the graphene mass density and $\alpha = 4 \times 10^{-7}$m$^{2}$s$^{-1}$.\cite{verma_13}


\subsection{Calculation of $\kappa(\omega, T)$}
\label{subsec:calculation}
As discussed earlier, the thermal conductivity can be computed using Eq.(\ref{eqn:thermal_conductivity}) and (\ref{eqn:thermal_memoryfunction}). Thus with the definitions of the thermal current (Eq.(\ref{eqn:thermal_current})) and the model Hamiltonian (Eq.(\ref{eqn:hamiltonian})), 
the imaginary part of the thermal memory function or the thermal scattering rate can be expressed as
\bea \nonumber
M''_{QQ}(\omega, T) &=& \frac{4\pi}{\chi_{QQ}^{0}(T) m^2} \sum_{\textbf{k} \textbf{k}'} \left[\left( \textbf{k} (\epsilon_{\textbf{k}} - \mu) - \textbf{k}'(\epsilon_{\textbf{k}'} - \mu) \right).\hat{n}\right]^{2} \\ \nonumber
&& \vert D(\textbf{k}-\textbf{k}') \vert^{2} (1-f_{\textbf{k}})f_{\textbf{k}'} n \\ \nonumber
&& \left\lbrace \frac{e^{\omega/T}-1}{\omega} \delta(\epsilon_{\textbf{k}}-\epsilon_{\textbf{k}'} - \omega_{\textbf{k}-\textbf{k}'} + \omega) \right. \\
&& \left. + (\text{terms with $\omega \rightarrow -\omega$}) \right\rbrace.
\label{eqn: memoryphonon}
\eea
Here $f_{\textbf{k}} = \frac{1}{e^{\beta(\epsilon_{\textbf{k}}-\mu)}+1}$ and $n = \frac{1}{e^{\beta \omega_{\textbf{q}}}-1}$ are the Fermi and the Boson distribution functions, $\beta$ is the inverse of the temperature and factor $4$ is for the two spin and two valley degeneracies.

To simplify Eq.(\ref{eqn: memoryphonon}), we convert the summations over momentum indices into the two dimensional energy integrals using the linear electron energy dispersion relation $\epsilon_{\textbf{k}} = v_{\text{F}}k$ and $\epsilon_{\textbf{k}'} = v_{\text{F}}k'$, where $v_{\text{F}}$ is the Fermi velocity. This linear dispersion relation distinguish the characteristics of the graphene from those of the three dimensional normal metals which follows the quadratic dispersion relation. Further these simplifications along with  
integrations over the angular parts yield 
\bea \nonumber
M''_{QQ}(\omega,T) &=& \frac{\epsilon_{F}^2 D_{0}^2}{4\pi^2 m^2 \rho_{m}   v_{F}^4 k_{F} \chi_{QQ}^{0}(T)} \int d\epsilon_{\textbf{k}} \int_{0}^{\lambda} dq \\ \nonumber
&& \frac{q^2}{\omega_{q}}\left( \omega_{q}^2 k_{F}^2 + (\epsilon_{\textbf{k}} - \mu)^2 q^2 + \frac{\omega_{q}(\epsilon_{\textbf{k}}-\mu)}{2} q^2 \right)\\ \nonumber
&& \sqrt{1-\left(\frac{q}{2k_{F}} \right)^{2}} (1-f(\epsilon_{\textbf{k}})) n \\ \nonumber
&&\left\lbrace \frac{e^{\omega/T}-1}{\omega} f(\epsilon_{\textbf{k}}- \omega_{q} + \omega) \right. \\
&& \left. + (\text{terms with $\omega \rightarrow -\omega$}) \right\rbrace. 
\label{eqn:imagmemphoacoustic}
\eea
Here we use the expression for the electron-phonon matrix element given in Eq.(\ref{eqn:electron-phononmatrixelement}) and the symbol $\lambda$ corresponds to the upper cut off value of the phonon momentum. Since in normal metals, the Fermi sphere is very large as compared to the Debye sphere, the phonons residing in the Debye sphere participate in scattering events and we restrict $\lambda$ to $q_{D}$, $q_{D}$ being the Debye momentum. While in the case of graphene, this does not remain same due to the smaller Fermi sphere than the Debye sphere. This allows only phonons residing below the Fermi surface to participate in the scattering phenomenon, hence restrict the upper cut off value of $q$ integral to $2k_{F}$. Further the above Eq.(\ref{eqn:imagmemphoacoustic}) for graphene can be solved for various acoustic phonons in the following subsections.

\subsubsection{Longitudinal/Transverse Acoustic Phonons (LA/TA)}
To compute $M_{QQ}(z,T)$ for the Longitudinal and the Transverse acoustic phonons (having linear dispersion relation), we define few dimensionless quantities such as $\frac{\epsilon_{\textbf{k}}-\mu}{T} = \eta$, $\frac{\omega_{q}}{T} = z$ and $\frac{\omega}{T} = x$, where $\omega_{q} = v_{\text{s}} q$, $v_{\text{s}} \equiv (v_{\text{LA}}, v_{\text{TA}})$. Using these variables and then performing the integral over the energy, Eq.(\ref{eqn:imagmemphoacoustic}) becomes
\bea \nonumber
M''_{QQ}(\omega,T) &=& \frac{\epsilon_{F}^2 D_{0}^2}{4\pi^2 m^2 \rho_{m} v_{F}^4 v_{s}^{5} k_{F}} \frac{T^{6}}{\chi_{QQ}^{0}(T)}\\ \nonumber
&& \int_{0}^{\Theta_{\text{BG}}/T} dz \frac{z^3}{e^{z}-1} \left( 1- \frac{z^2 T^2}{2\Theta_{\text{BG}}^{2}}\right) \\ \nonumber
&& \left\lbrace \frac{x-z}{e^{x-z}-1} \frac{e^{x}-1}{x} \left(  \frac{\Theta_{\text{BG}}^{2}}{4T^2} + \frac{\pi^2}{3} + \frac{(x-z)^2}{3} \right.\right. \\ 
&& \left.\left.+ \frac{z(x-z)}{4} \right) + (\text{terms with $\omega \rightarrow - \omega$}) \right\rbrace.
\label{eqn:LA_TA_final_MF}
\eea
Here $\Theta_{\text{BG}}$ is the Bloch-Gr\"ueinsen temperature and is equal to $2k_{F}v_{s}$. Further, the above expression in different frequency and temperature domains can be discussed or analyzed as follows:\\
\textbf{Case-I: The zero frequency limit i.e. $\omega \rightarrow 0$}\\
In this limit in Eq.(\ref{eqn:LA_TA_final_MF}), $M''_{Q}(T)$ becomes
\bea \nonumber
M''_{QQ}(T) &=& \frac{\epsilon_{F}^2 D_{0}^2}{2\pi^2 m^2 \rho_{m} A v_{F}^4 v_{s}^{5} k_{F}} \frac{T^{6}}{\chi_{QQ}^{0}(T)}\\ \nonumber
&& \int_{0}^{\Theta_{\text{BG}}/T} dz \frac{z^4}{(e^{z}-1)^2} \left( 1- \frac{z^2 T^2}{2\Theta_{\text{BG}}^{2}}\right) \\
&& \left(\frac{\Theta_{\text{BG}}^{2}}{4T^{2}} + \frac{\pi^2}{3} + \frac{z^{2}}{12}  \right).
\label{eqn:LA_TA_memory_dc}
\eea
Here, we find that $M''_{QQ}(\omega,T)$ for the case of interaction of the electrons with the longitudinal or transverse phonons leads to the linear and the quadratic temperature dependence in the high ($T \gg \Theta_{\text{BG}}$) and low ($T \ll \Theta_{\text{BG}}$) temperature regimes respectively.\\
Now the thermal conductivity Eq.(\ref{eqn:thermal_conductivity}) in the zero frequency limit can be written as\cite{bhalla_16}
\bea
\kappa(T) &=& \frac{1}{T}\frac{\chi_{QQ}^{0}(T)}{M''_{QQ}(T)} \approx \frac{T}{M''_{QQ}(T)}.
\label{eqn:thermal_dc_actual}
\eea
Thus the thermal conductivity depends inversely on the thermal memory function. From Eqs.(\ref{eqn:LA_TA_memory_dc}) and (\ref{eqn:thermal_dc_actual}), we find that $\kappa(T)$ for the case of LA and TA phonons varies inversely with the temperature and becomes saturate at the low and the high temperature regimes (as shown in Table \ref{tab:phonontable}). These are in accord with the results existed in the literature\cite{munoz_12, kim_16}.\\
\textbf{Case-II: Finite frequency regimes}\\
In finite frequency regimes, the asymptotic results of the thermal memory function and the corresponding thermal conductivity in different temperature and the frequency regimes are shown in Table \ref{tab:phonontable}. Here, we observe that $M''_{QQ}(\omega,T)$ shows frequency independent behavior at extremely low frequency (or dc limit) and then in the intermediate regimes, the complicated behavior is observed. In the high frequency regime, due to more excitations, it varies quadratically with the increase in the frequency.\\
Now from Eq.(\ref{eqn:thermal_conductivity}), the real part of the thermal conductivity is expressed as
\bea
\text{Re}[\kappa(\omega,T)] &=& \frac{\chi_{QQ}^{0}(T)}{T} \frac{ M''_{QQ}(\omega,T)}{\omega^2 + (M''_{QQ}(\omega,T))^2},
\label{eqn:thermal_ac}
\eea
where $M''_{QQ}(\omega,T)$ for different regimes are given in Table \ref{tab:phonontable}.
\begin{widetext}
\begin{table}[htb]
\caption{The results of thermal memory function and the thermal conductivity for the interaction of electrons with LA/TA and ZA phonons in different frequency and temperature domains.}
\begin{center}
 \begin{tabular}{|>{\centering\arraybackslash}m{1in}| >{\centering\arraybackslash}m{1.5in}| >{\centering\arraybackslash}m{1.5in}| >{\centering\arraybackslash}m{1.5in}| >{\centering\arraybackslash}m{1.5in}|} 
 \hline
 \multirow{2}{*}{Regimes} & \multicolumn{2}{c}{LA/TA phonons} & \multicolumn{2}{c|}{ZA phonons} \\ 
   \cline{2-5}&Thermal memory function \newline $1/\tau_{\text{th}}$ or $M''_{QQ}$ & Thermal conductivity, \newline $\kappa$ &Thermal memory function \newline $1/\tau_{\text{th}}$ or $M''_{QQ}$ & Thermal conductivity, \newline $\kappa$ \\
   \hline
   $\omega = 0$, $T \gg \Theta_{\text{BG}}$ &$T^{1}$ & $T^{0}$& $T^{1}$& $T^{0}$\\ [0ex]
   \hline
   $\omega = 0$, $T \ll \Theta_{\text{BG}}$ &$T^{2}$ & $T^{-1}$& $T^{1/2}$& $T^{1/2}$\\ [0ex]
   \hline
   $\omega \ll T \ll \Theta_{\text{BG}}$ &$T^{2}$ & $T^{3} \omega^{-2}$& $T^{1/2}$& $T^{3/2} \omega^{-2}$\\ [0ex]
   \hline
   $\omega \ll \Theta_{\text{BG}} \ll T$ &$T^{1}$ &$T^{2} \omega^{-2}$ & $T^{1}$& $T^{2} \omega^{-2}$\\ [0ex]
   \hline
   $T \ll \omega \ll \Theta_{\text{BG}}$ &$T^{3} \omega^{-1} e^{\omega/T}$ & $T^{4} \omega^{-3} e^{\omega/T}$& $T^{5/2} \omega^{-1} e^{\omega/T}$& $T^{7/2} \omega^{-3} e^{\omega/T}$\\ [0ex]
   \hline
   $\Theta_{\text{BG}} \ll \omega \ll T$ &$T^{1}$ & $T^{2} \omega^{-2}$& $T^{1}$ & $T^{2} \omega^{-2}$\\ [0ex]
   \hline
   $\Theta_{\text{BG}} \ll T \ll \omega$ &$T^{-1} \omega^{2}$ & $T^{0} \omega^{0}$ & $T^{-1} \omega^{2}$ & $T^{0}\omega^{0}$\\ [0ex]
   \hline
    $T \ll \Theta_{\text{BG}} \ll \omega$ &$T^{2}\omega^{2}$ & $T^{3}$& $\omega^{2}T^{-1/2}$& $T^{1/2}$\\ [0ex]
   \hline
  \end{tabular}
  \end{center}
\label{tab:phonontable}
\end{table}
\end{widetext}
In the perturbative regime of small electron-phonon couplings, we assume that the frequency dependent thermal memory function is small. Using this assumption, Eq.(\ref{eqn:thermal_ac}) can be written as\cite{bhalla_16, bhalla_16a}
\bea \nonumber
\text{Re}[\kappa(\omega,T)] &\approx& \frac{\chi_{QQ}^{0}(T) }{T}\frac{M''_{QQ}(\omega,T)}{\omega^2} \approx \frac{T M''_{QQ}(\omega,T)}{\omega^2}.\\
\eea
Here we use the temperature variation of the static correlation function. On substituting the variation of the temperature and the frequency dependent thermal memory function, we conclude that the thermal conductivity at high frequency shows frequency independent behavior. While at the low frequency, it gives large conductivity due to the weakly frequency dependent behavior of the thermal memory function. These behaviors are summarized in Table \ref{tab:phonontable}.
\subsubsection{Flexural Acoustic Phonons (ZA)}
Now, in the case of the flexural acoustic phonons having quadratic dispersion\cite{amorim_13} i.e. $\omega_{q} = \alpha q^{2}$, the thermal memory function can be computed in a similar fashion as done in the case of the LA/TA phonons.\\
Following the same procedure, the Eq.(\ref{eqn:imagmemphoacoustic}) for ZA phonons is written as
\bea \nonumber
M''_{QQ}(\omega,T) &=& \frac{\epsilon_{F}^2 D_{0}^2 }{8\pi^2 m^2 \rho_{m} v_{F}^4 \alpha^{3/2} k_{F}} \frac{T^{7/2}}{\chi_{QQ}^{0}(T)}\\ \nonumber
&& \int_{0}^{\Theta_{\text{BG}}/T} dz \frac{z^{1/2}}{e^{z}-1} \left( 1- \frac{z^2 T^2}{2\Theta_{\text{BG}}^{2}}\right) \\ \nonumber
&& \left\lbrace \frac{x-z}{e^{x-z}-1} \frac{e^{x}-1}{x} \left(  \frac{\Theta_{\text{BG}}}{4T} + \frac{\pi^2}{3} + \frac{(x-z)^2}{3} \right.\right. \\ 
&& \left.\left.+ \frac{z(x-z)}{4} \right) + (\text{terms with $\omega \rightarrow - \omega$}) \right\rbrace.
\label{eqn:ZA_final_MF}
\eea
This is analyzed in different frequency and temperature domains as follows.\\
\textbf{Case-I: The zero frequency limit i.e. $\omega \rightarrow 0$}\\
Using this limit in Eq.(\ref{eqn:ZA_final_MF}), we have
\bea \nonumber
M''_{QQ}(\omega,T) &=& \frac{\epsilon_{F}^2 D_{0}^2 }{4\pi^2 m^2 \rho_{m} v_{F}^4 \alpha^{3/2} k_{F}} \frac{T^{7/2}}{\chi_{QQ}^{0}(T) }\\ \nonumber
&& \int_{0}^{\Theta_{\text{BG}}/T} dz \frac{z^{3/2} e^{z}}{(e^{z}-1)^{2}} \left( 1- \frac{z^2 T^2}{2\Theta_{\text{BG}}^{2}}\right) \\ 
&& \left(  \frac{\Theta_{\text{BG}}}{4T} + \frac{\pi^2}{3} + \frac{z^2}{12} \right).
\label{eqn:ZA_memory_dc}
\eea
Further in the high and the low temperature regimes, Eq.(\ref{eqn:ZA_memory_dc}) shows that the thermal memory function $M''_{QQ}(T)$ varies as a square root and linearly with temperature at $T \gg \Theta_{\text{BG}}$ and $T \ll \Theta_{\text{BG}}$ (shown in Table \ref{tab:phonontable}). Accordingly, the thermal conductivity (Eq.(\ref{eqn:thermal_dc_actual})) leads to the $T^{1/2}$ power law behavior at the low temperature and temperature independent behavior in the high temperature.\\
\textbf{Case-II: Finite frequency regimes}\\
In this case, we have shown the asymptotic results in Table \ref{tab:phonontable}. It is observed that the frequency variation for $M''_{QQ}(\omega,T)$ and the corresponding $\kappa(\omega,T)$ is same as the case for the longitudinal or the transverse phonons. But the temperature variations are different. At the temperature higher than the BG temperature, the ZA phonons show identical temperature dependent behavior as the LA/TA phonons. On the other hand, at the low temperature then the BG temperature, the temperature dynamics of ZA phonons is different from the LA/TA phonons.

\section{Results}
\label{sec:results}
In this section, we present our findings for the thermal scattering rate and the thermal conductivity for different cases.
\figphononA
\subsection{Thermal Conductivity in zero frequency limit}
In Fig. \ref{fig:memory_LATA}, $M''_{QQ}(T)$ is plotted as a function of $T$ for LA and TA phonons at different $\Theta_{\text{BG}}$ which depends on the carrier density $n$. Here we separately plot it by setting $\Theta_{\text{BG}} \approx 57 \sqrt{n}$ and $\Theta_{\text{BG}} \approx 38 \sqrt{n}$ for the LA and the TA phonons respectively. Also, we have scaled the $M''_{QQ}(\omega,T)$ with $M''_{0}$($= \frac{6 \epsilon_{F}^2 D_{0}^2}{\pi^3 \rho_{m} v_{F}^3 v_{s}^{5} k_{F}}$) It is observed that the thermal memory function increases linearly with increase in the temperature in the high temperature $T\gg\Theta_{\text{BG}}$ and non linearly in the low temperature $T\ll\Theta_{\text{BG}}$ regimes. Also, it decreases with the increase in the carrier density or $\Theta_{\text{BG}}$. This decrease is due to the linear density of states which provides more phase space to phonons to scatter. This results in the less electron-phonon scattering rate. On comparing the Fig. \ref{fig:memory_LA} and \ref{fig:memory_TA}, it is found that the magnitude of the thermal memory function for the TA phonons is more than the LA phonons. This is due to the low phonon velocity of the TA phonons.

\figphononB
The corresponding thermal conductivity for LA and TA phonons is shown in Fig. \ref{fig:thermal_LATA}. Here for $T \ll \Theta_{\text{BG}}$, the thermal conductivity reduces with the increase in $T$ and at $T \gg \Theta_{\text{BG}}$, it saturates. These observed features are in accord with the results existed in the literature\cite{munoz_12, kim_16}. In the intermediate regime i.e. around $\Theta_{\text{BG}}$, the small dip is observed which is due to the consideration of the normal process scattering in the system.

\figphononE
For the flexural (ZA) phonons, $M''_{QQ}(T)$ is shown with the variation in the temperature in Fig. \ref{fig:thermal_memory_ZA}. Here we have set $\Theta_{\text{BG}} \approx 0.1 n$. This small $\Theta_{\text{BG}}$ ensures that these phonons play significant role in the low temperature behavior of the thermal conductivity of graphene. Here the value of $M''_{0}$ is $\frac{3\epsilon_{F}^{2} D_{0}^{2}}{\pi^{3} \rho_{m} v_{F}^{3} k_{F}^{4} \alpha^{5/2}}$. It is observed that the thermal memory function increases with the increase in the temperature by power law $T^{1/2}$ which further results the increase in the thermal conductivity as $T^{1/2}$ law. But at the high temperature, it increases linearly similar to the case of LA/TA phonons and hence results in the temperature independent thermal conductivity.\\ 

\subsection{Thermal Conductivity in finite frequency regime}
To discuss our results at the finite frequency, we plot $M''_{QQ}(\omega,T)/M''_{0}$ and $\text{Re}[\kappa(\omega,T)]/\kappa_{0}$ with the variation in the frequency at different temperature ratio i.e. $T/\Theta_{\text{BG}}$.\\ 

\figphononC

In Fig. \ref{fig:memory_ac_LATA} and \ref{fig:memory_ac_ZA}, we find that in the high frequency regime, the thermal memory function increases with the increase in the frequency. While at the low frequency, it shows saturation behavior. Next, using this variation, we plot the real part of the thermal conductivity (\ref{eqn:thermal_ac}) in Fig. \ref{fig:thermal_ac_LA}, \ref{fig:thermal_ac_TA} and \ref{fig:thermal_ac_ZA}. From the frequency behavior of the thermal conductivity, we observe that it is suppressed by the factor $1/\omega^{2}$ in the high frequency regime. While in the low frequency regime, the frequency variations are observed. These frequency dependent behavior of $\kappa(\omega,T)$ is identical to the case of the metal. This gives the signature that the two dimensional scenario modifies the temperature variation of the thermal conductivity and does not give effect on the frequency variation of $\kappa(\omega,T)$.\\
\figphononD

\figphononF

\figphononG
By comparing Fig. \ref{fig:thermal_LA},\ref{fig:thermal_TA} and \ref{fig:thermal_ZA}, we note that the magnitude of thermal conductivity $\kappa(T)$ is different in all three cases. This is due to the different values such as $v_{\text{LA}} = 21.2 \times 10^{-3}$ ms$^{-1}$, $v_{\text{TA}} = 14.1 \times 10^{-3}$ ms$^{-1}$ and $\alpha = 4.7 \times 10^{-7}$ m$^{2}$s$^{-1}$ for the longitudinal, transverse and the flexural phonons respectively. Due to it, the LA phonons contribute more to the thermal conductivity as compared to the TA and ZA phonons. This can also be explained as follows.

In the total thermal conductivity, $\kappa^{-1}(T) = \kappa_{\text{LA}}^{-1}(T) + \kappa_{\text{TA}}^{-1}(T) + \kappa_{\text{ZA}}^{-1}(T)$. This shows that at the high temperature $T \gg \Theta_{\text{BG}}$, $\kappa^{-1}(T) \approx \text{constant}$ and at the low temperature i.e. $T \ll \Theta_{\text{BG}}$, $\kappa^{-1}(T) \approx B\left( \frac{T}{v_{\text{LA}}^{5}} + \frac{T}{v_{\text{TA}}^{5}}  + \frac{T^{-1/2}}{\alpha^{5/2}} \right)$. Here, we find that  at the low temperature, the contribution of the LA phonons is more than others. And the total thermal conductivity decreases approximately linearly with the temperature. Because of the small value of the magnitude of thermal conductivity of ZA phonons, it does not effect much to the total thermal conductivity.\\
\section{Conclusion}
\label{sec:conclusion}
Graphene is an unique system as it has two dimensional nature, unusual electron dispersion relation, etc. which make its properties different from what is found from the normal three dimensional metal. In the later case, electrons follow the quadratic energy dispersion relation while in graphene, these follow the linear energy dispersion relation. It also shows unusual phonon modes that do not exist in normal metals. In some sense, these characteristics make the study of graphene novel and more promising.

In the present study, the effect of the electron-acoustic phonon interactions to the electronic thermal conductivity is analyzed in detail. These analytic calculations for $\kappa(\omega,T)$ have been performed by using memory function formalism which is beyond the relaxation time approximation. We find that the electronic thermal conductivity for various acoustic phonons shows different power law behavior due to the linear and quadratic phonon dispersion relations. These power law predictions are in agreement with the existing results in the literature\cite{munoz_12, kim_16}. It is also showed that in the total thermal conductivity, the contribution of the ZA phonons is extremely small.

For the finite frequency cases, we have studied the dynamics of the thermal conductivity due to the electron-phonon interaction which is  identical to the case of three dimensional system such as metal\cite{bhalla_16}. But due to the semi-metallic character of the graphene, it's dynamical study may give information about the heat control for the reliable use of electronic devices.
\section*{Acknowledgement}
The author is thankful to N. Singh, N. Das, S. K. Haldar and A. Atreya for helpful suggestions and discussions.

\end{document}